\documentclass[prb,floatfix,twocolumn,showpacs,amsmath,amssymb]{revtex4-1}
\usepackage{graphicx}
\usepackage{epstopdf}
\usepackage{epsfig}
\usepackage{dcolumn}
\usepackage{bm}

\newcommand{\dagga}{{\phantom{\dagger}}}

\begin{document}

\title{Variational Monte Carlo approach to the two-dimensional Kondo lattice 
model}
\author{Mohammad Zhian Asadzadeh, Federico Becca, and Michele Fabrizio}
\affiliation{Democritos Simulation Center CNR-IOM Istituto Officina dei 
Materiali and International School for Advanced Studies (SISSA), Via Bonomea 
265, 34136 Trieste, Italy}

\date{\today}

\begin{abstract}
We study the phase diagram of the Kondo-lattice model with nearest-neighbor 
hopping in the square lattice by means of the variational Monte Carlo 
technique. Specifically, we analyze a wide class of variational wave functions
that allow magnetic and superconducting order parameters, so to assess the 
possibility that superconductivity might emerge close to the magnetic 
instability, as often observed in heavy fermion systems. Indeed, we do find 
evidence of $d$-wave superconductivity in the paramagnetic sector, i.e., when
magnetic order is not allowed in the variational wave function. However, when
magnetism is allowed, it completely covers the superconducting region, which 
thus disappears from the phase diagram. 
\end{abstract}


\maketitle

\section{Introduction}\label{sec:intro}

The Kondo lattice model (KLM) describes localized magnetic moments that 
interact with a single band of itinerant electrons via an antiferromagnetic 
exchange coupling $J$. This model has been introduced long ago by 
Doniach,~\cite{doniach} and, since then, has been widely invoked to describe 
the physics of heavy-fermion materials.~\cite{stewart} Three ingredients 
define the KLM: the bandwidth $W$ of the conduction electrons (usually denoted
by $c$ electrons), their density $n_c$, and the Kondo exchange $J$ that 
controls the coupling between the local moments and the spin density of 
$c$-electrons. The localized moments (denoted by $f$ electrons) are anchored 
to the sites of a regular lattice. The phase diagram of the KLM depends in a 
non-trivial way on $J/W$ and the electron density $n_c$. In one spatial 
dimension, the KLM has been intensively studied and shows three distinct 
phases.~\cite{tsunetsugu} In the compensated case, one conduction electron 
per localized spin, i.e., $n_c=1$, it is a so-called Kondo insulator with a 
charge as well as a spin gap without any magnetic order, a kind of spin-liquid
insulator. For $n_c \not=1$, it is either a paramagnetic metal for low $J/W$,
or a ferromagnetic metal for larger $J/W$. In higher dimensions, where the
$SU(2)$ spin symmetry can be spontaneously broken, the phase diagram is 
expected to enrich and display a critical point separating a paramagnetic 
heavy-fermion metal from a different metallic phase with long-range magnetic 
order.~\cite{doniach,lacroix} Indeed, in analogy with the Kondo effect that 
occurs in the case of a single magnetic impurity, the conduction electrons 
may screen the localized moments, thus forming a global singlet state. However,
such a Kondo screening is thwarted by the tendency to magnetic ordering of the
localized moments. In fact, the latter ones interact mutually via the 
Ruderman-Kittel-Kasuya-Yosida (RKKY) mechanism, through an effective exchange 
$J_{eff}(q) \propto -J^2 \, \Re e\chi(q,\omega=0)$ mediated by the magnetic 
polarization $\chi(q,\omega)$ of the conduction band. The competition between 
Kondo screening (that favors a paramagnetic ground state) and RKKY interaction
(that favors magnetically ordered states) is the heart of a frustrating 
behavior, which may lead to genuine quantum phase transitions.\cite{doniach} 
Furthermore, other physical processes may profit from the balanced competition
between Kondo screening and magnetic ordering nearby the transition, and make 
novel phases to intrude, most notably superconductivity. Indeed, the discovery
of superconductivity in CeCu$_2$Si$_2$,~\cite{steglich} and subsequently in 
several other heavy-fermion materials, unveiled the rich variety of phenomena 
of these strongly-correlated systems. It is widely believed that, in 
heavy-fermion compounds, superconductivity is not the conventional 
phonon-mediated one, but most likely it is caused by antiferromagnetic spin 
fluctuations.~\cite{mathur}

Recent experiments of the Hall coefficient in the heavy-fermion material 
YbRh$_2$Si$_2$,~\cite{pashen} have provided new intriguing elements that 
renewed interest in the phase diagram of the KLM. In particular, the rapid jump
of the Hall coefficient~\cite{pashen,schoeder,gegenwart,custers,coleman,friedmann,custers2} 
is suggestive of a sudden change of the Fermi surface topology at, or close 
to, the quantum phase transition between the non-magnetic metal and the 
magnetic one. In the conventional view of heavy-fermions, the localized spins 
are promoted in the conduction band through the Kondo effect. A mass-enhanced 
Fermi liquid is thus settled down, with a ``large'' Fermi surface that 
includes the conduction as well as the localized electrons. In the simplest 
scenario of a spin-density-wave quantum phase transition,~\cite{hertz,millis} 
magnetic ordering is indeed expected to modify the topology of the Fermi 
surface by appearance of the magnetic Bragg reflections. In addition, the 
Fermi surface reconstruction across the magnetic field induced transition in 
YbRh$_2$Si$_2$ can also be simply interpreted as a Zeeman driven Lifshitz 
transition of the heavy quasiparticles.~\cite{vojta} However, an alternative 
scenario is possible in which the Kondo effect dies out at the transition 
point,~\cite{si} hence the local moments suddenly stop contributing to the 
volume of the Fermi surface, which then counts only the number of $c$ 
electrons. In the language of the Anderson lattice model, which maps for 
strong repulsion onto the KLM, the death of the Kondo effect would translate 
into the Mott localization of the $f$ electrons, whose magnetic ordering would
then be only a by-product rather then the driving source. Indeed, there are 
suggestions that the Kondo-breakdown and the onset of magnetic order in the 
Anderson lattice model are distinct phenomena, which may occur at different 
points of the phase diagram.~\cite{senthil,deleo} This scenario has been 
indirectly supported by variational Monte Carlo calculations~\cite{ogata}
and by the Gutzwiller approximation~\cite{lanata} in the KLM on a square 
lattice. Indeed, these works found evidence of two distinct phase transitions,
one given by the continuous appearance of long-range magnetic order and another
one related to an abrupt topological change of the Fermi surface. The same 
outcome of two distinct transitions has been later observed also within the 
dynamical cluster approximation (DCA).~\cite{assaad} In spite of all efforts, 
such an interesting issue like remains open.  

In this paper, we investigate the KLM in two dimensions paying attention not 
only to magnetism, but also to the possible emergence of superconductivity in 
its proximity. We use both mean-field and variational Monte Carlo approaches. 
As far as the former one is concerned, an analytical treatment of the 
long-range antiferromagnetic phase is possible only at compensation, $n_c=1$, 
where calculations have been already performed.~\cite{zhang} Here, we go 
beyond the results of Ref.~\onlinecite{zhang} and consider also the 
uncompensated regime, $n_c<1$, by solving numerically the Hartree-Fock 
equations. Furthermore, we generalize the previous calculations based upon 
the variational Monte Carlo approach~\cite{ogata} or the Gutzwiller 
approximation~\cite{lanata} and introduce additional correlations inside the 
trial wave function, among which superconducting ones. In particular, our 
variational calculations show that superconductivity is indeed stabilized in 
the paramagnetic sector in a region of the phase diagram close to $n_c=1$ and 
not too large $J/W$. However, when magnetism is allowed in the variational 
wave function, an antiferromagnetic phase emerges and completely covers the 
superconducting dome. Therefore, at least within the variational approach and 
in our model where the only source of magnetic frustration is deviation from 
the compensated regime, we do not find any superconducting region at the 
border between antiferromagnetic and heavy-fermion metal.  

The paper is organized as follow: in section~\ref{sec:model}, we introduce 
the microscopic model and discuss the methods that we use; 
in section~\ref{sec:results}, we present our numerical results for both 
mean-field and variational approximations; and, finally, 
in section~\ref{sec:conc}, we draw our conclusions.

\section{Model and methods}\label{sec:model}

The KLM model on the two-dimensional square lattice is defined by:
\begin{equation}\label{eq:KLM}
{\cal H} = -t \sum_{\langle i,j\rangle,\sigma} 
c^\dag_{i,\sigma} c^\dagga_{j,\sigma} + h.c. + J \sum_{i} {\bf S}_i \cdot {\bf s}_i 
\end{equation}
where $\langle i,j \rangle$ denotes nearest-neighbor sites $i$ and $j$ and 
$c^\dag_{i,\sigma}$ ($c^\dagga_{i,\sigma}$) creates (destroys) an itinerant 
electron at site $i$ with spin $\sigma$; ${\bf s}_i=(s^x_i,s^y_i,s^z_i)$ is 
the spin operator for the $c$-electrons, i.e., 
$s^\alpha_i=1/2 \sum_{\sigma,\sigma^\prime} c^\dag_{i,\sigma} 
\tau^\alpha_{\sigma,\sigma^\prime} c^\dagga_{i,\sigma^\prime}$, $\tau^\alpha$ 
being the Pauli matrices. Similarly, ${\bf S}_i=(S^x_i,S^y_i,S^z_i)$ is the 
spin operator for the localized $f$ electrons, 
$S^\alpha_i=1/2 \sum_{\sigma,\sigma^\prime} f^\dag_{i,\sigma} 
\tau^\alpha_{\sigma,\sigma^\prime} f^\dagga_{i,\sigma^\prime}$. 
By constraint there is one $f$ electron per each site. The exchange coupling 
is antiferromagnetic, i.e., $J>0$, and we shall take all energies measured in 
units of $t$.

\subsection{Mean-field approach}\label{sec:meanfield}

The simplest approach to the KLM of Eq.~(\ref{eq:KLM}) is the mean-field
approximation, in which the spin-spin interaction is decoupled to bring about 
a non-interacting Hamiltonian. We shall implement Hartree-Fock by assuming 
finite the following  average values, to be determined self-consistently:
\begin{eqnarray}
V &=& \langle c^\dag_{i,\sigma} f^\dagga_{i,\sigma}\rangle=
\langle f^\dag_{i,\sigma} c^\dagga_{i,\sigma}\rangle, \label{V-MF}\\
m_c &=& -(-1)^{R_i} \, \langle s^z_i \rangle, \label{mc-MF}\\
m_f &=& (-1)^{R_i}\, \langle S^z_i \rangle.\label{mf-MF}
\end{eqnarray}
$V$ denotes the (site-independent) $c{-}f$ hybridization, responsible in 
mean-field for the creation of the Kondo singlet; $m_c$ and $m_f$ are the 
staggered magnetizations of conduction and localized electrons, respectively,
which have opposite sign because of the antiferromagnetic exchange $J$.
In addition, a Lagrange multiplier $\mu_f$ must be included to enforce 
(on average) the $f$-orbital occupancy, i.e., $n_f=1$.

In momentum space, the antiferromagnetic mean-field Hamiltonian can be 
cast in a $4 \times 4$ matrix form:
\begin{equation}
\begin{split}
{\cal H}_{\rm AF} =\sum\limits_{k\in MBZ,\sigma}
\left[
\begin{array}{cccc}  
c^\dag_{k,\sigma} & c^\dag_{k+Q,\sigma} & f^\dag_{k,\sigma} & f^\dag_{k+Q,\sigma}
\end{array} 
\right] 
\times \\
\begin{bmatrix}
 \epsilon_{k}             & \frac{1}{2} J m_f \sigma & -\frac{3}{4} J V          & 0                         \\ 
 \frac{1}{2} J m_f \sigma & -\epsilon_k              & 0                         & -\frac{3}{4} J V          \\
 -\frac{3}{4} J V         & 0                        & -\mu_f                    & -\frac{1}{2} J m_c \sigma \\
 0                        & -\frac{3}{4} J V         & -\frac{1}{2} J m_c \sigma & -\mu_f
\end{bmatrix}
\left[ 
\begin{array}{c} 
c^\dagga_{k,\sigma}   \\ 
c^\dagga_{k+Q,\sigma} \\ 
f^\dagga_{k,\sigma}   \\ 
f^\dagga_{k+Q,\sigma} 
\end{array} 
\right],
\end{split}
\end{equation}
where the sum over $k$ is restricted to the reduced (magnetic) Brillouin zone.
In order to compute the total energy, the constant term
$(3JV^2/2+Jm_cm_f+\mu_f)N$ must be added ($N$ being the number of sites).

The paramagnetic state is found by imposing $m_c=m_f=0$ and corresponds to
the $2 \times 2$ matrix Hamiltonian:
\begin{equation}
{\cal H}_{\rm PM}=\sum\limits_{k,\sigma}
\left[ 
\begin{array}{cc}  
c^\dag_{k,\sigma} & f^\dag_{k,\sigma}
\end{array} 
\right]
\begin{bmatrix} 
 \epsilon_{k} & V     \\ 
 V            & -\mu_f
\end{bmatrix}
\left[ 
\begin{array}{c} 
c^\dagga_{k,\sigma} \\ 
f^\dagga_{k,\sigma}
\end{array} 
\right].
\end{equation}

The self-consistency conditions Eqs.~(\ref{V-MF}),~(\ref{mc-MF}), 
and~(\ref{mf-MF}) are solved numerically on finite size systems with $N$ sites,
number that we scale to get reliable estimates in the thermodynamic limit. 
We mention that an analytic solution of the problem is possible only in the 
compensated regime, while in general numerical calculations are needed. 
In practice, we numerically diagonalize the $4\times 4$ matrix for all $k$ 
points independently and then fill the bands with the lowest orbitals; 
the mean-field parameters are numerically calculated and the procedure is 
iterated until convergence is reached. 
At the mean-field level, a superconducting singlet order parameter 
$\sum_\sigma\,\sigma \,\langle c^\dag_{k,\sigma}f^\dag_{k,-\sigma}\rangle$
is not independent from the hybridization, because of the charge-isospin 
$SU(2)$ symmetry displayed by the $f$ electrons.  

\begin{figure}
\includegraphics[width=\columnwidth]{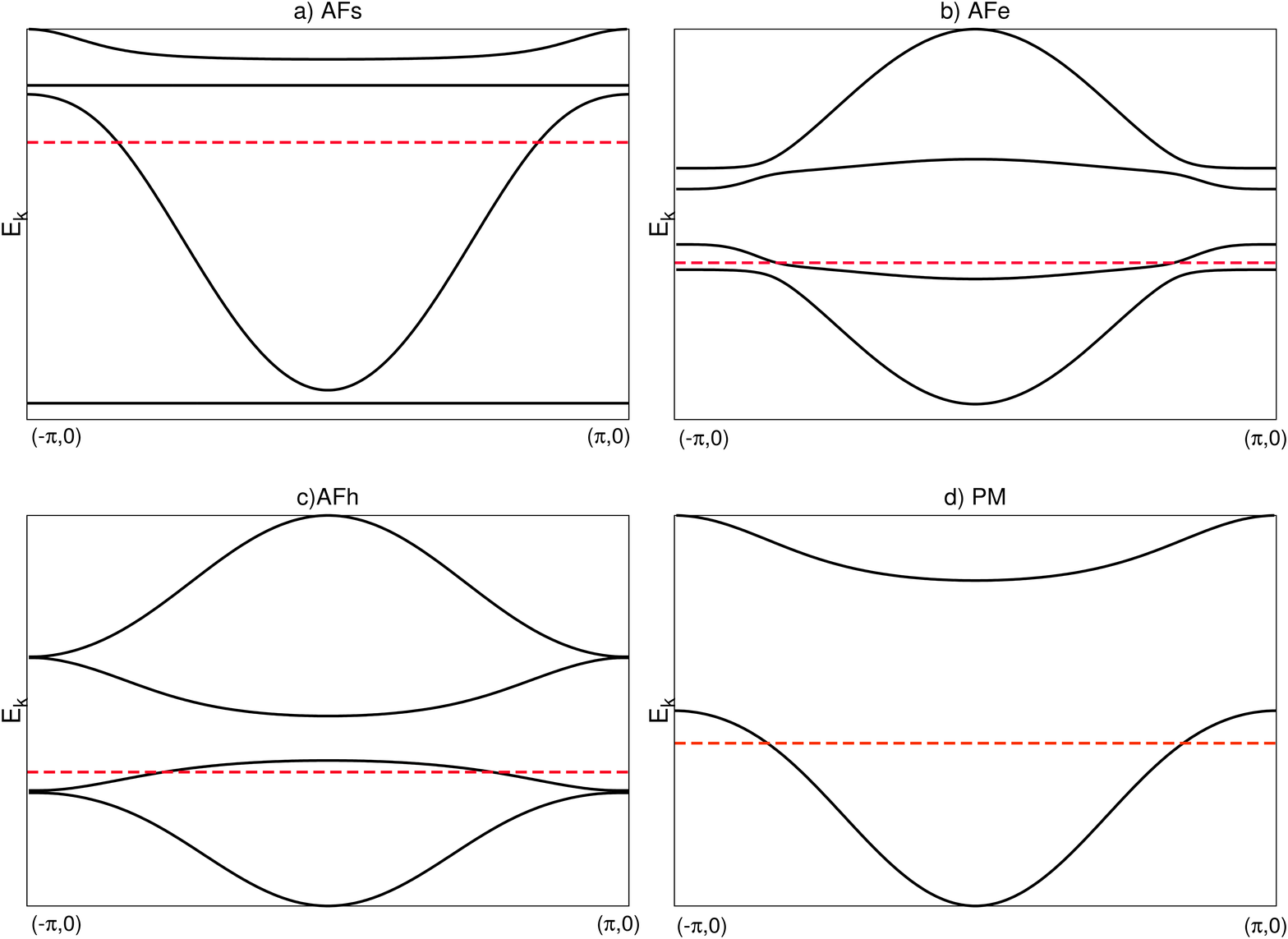}
\caption{\label{fig:bands}
(Color on-line) Quasi-particle band structure for various phases: the AFs (a) 
has flat $f$ bands, due to the full decoupling between $f$ and $c$ electrons 
and a small Fermi surface; the antiferromagnetic phases with electron-like 
and hole-like Fermi surfaces are denoted by AFe (b) and AFh (c), respectively;
finally, PM (d) is a paramagnetic (metallic) phase. The dashed red line 
indicates the chemical potential.}
\end{figure}

Within the AF state, we find three different cases, depending on the magnitude
of variational parameters. In the following, we will adopt the notations of 
Ref.~\onlinecite{ogata}. Whenever the hybridization parameter vanishes, i.e.,
$V=0$, the localized electrons decouple from the conducting ones and they do 
not contribute to the volume enclosed by the Fermi surface, in this case we 
have an antiferromagnetic state with a ``small'' Fermi surface (denoted by 
AFs). By adding a small hybridization to the AFs, we end up with a state which
has an electron-like Fermi surface, the so-called AFe. Finally, in the case 
where the hybridization $V$ is large and the magnetic order parameter small, 
we have a hole-like Fermi surface, the so-called AFh. Here, the $f$ electrons 
participate to the total volume enclosed by the Fermi surface, which is 
therefore ``large''. A qualitative picture of all these cases, together 
with the simple paramagnetic state is depicted in Fig.~\ref{fig:bands}.

\subsection{Variational wave functions}\label{sec:variational}

In order to go beyond the mean-field approximation, we consider correlated
variational wave functions, in which the constraint of one $f$ electron per 
site is imposed {\it exactly} via a Gutzwiller projector. This is achieved 
through the projected variational wave function:
\begin{equation}
|\Psi\rangle = {\cal P}_{f}|\Psi_{\rm MF}\rangle,
\end{equation}
where ${\cal P}_{f}$ is the projector which enforces single occupation of $f$ 
orbitals on each site. Here, $|\Psi_{MF}\rangle$ is an uncorrelated wave 
function defined as the ground state of a non-interacting variational 
Hamiltonian ${\cal H}_{\rm MF}$ that may contain, in addition to the 
mean-field parameters of the previous section~\ref{sec:meanfield}, also 
direct $f{-}f$ hopping as well as superconducting terms:
\begin{eqnarray}
\label{eq:chiff}
\chi_{i,j}^{ff}   &=& \langle f^\dag_{i,\sigma} f^\dagga_{j,\sigma} \rangle, \\
\label{eq:deltaff}
\Delta_{i,j}^{ff} &=& \langle f^\dag_{i,\uparrow} f^\dag_{j,\downarrow}
                   +f^\dag_{j,\uparrow} f^\dag_{i,\downarrow} \rangle, \\
\label{eq:deltacc}
\Delta_{i,j}^{cc} &=& \langle c^\dag_{i,\uparrow} c^\dag_{j,\downarrow}
                   +c^\dag_{j,\uparrow} c^\dag_{i,\downarrow} \rangle, \\
\label{eq:deltacf}
\Delta_{i,j}^{cf} &=& \langle c^\dag_{i,\uparrow} f^\dag_{j,\downarrow}
                   +f^\dag_{j,\uparrow} c^\dag_{i,\downarrow} \rangle,
\end{eqnarray}
in $s$-wave or $d$-wave configurations. An on-site $c{-}c$ pairing has been
also considered.

Therefore, in the following, we shall consider four kind of uncorrelated 
variational wave functions: (1) paramagnetic, (2) antiferromagnetic, 
(3) superconducting, and, finally, (4) with coexisting antiferromagnetism and 
superconductivity. The variational parameters of the non-interacting 
Hamiltonian are determined so as to minimize the total energy. Because of the 
presence of the Gutzwiller projector ${\cal P}_f$ we have to use a 
variational Monte Carlo technique~\cite{sorella} to compute the total energy.
In practice, we minimize the variational energy for all the previous states 
as a function of the exchange coupling $J$ and the electron density $n_c$. 
Calculations have been performed on clusters with $64$, $100$, $144$, and 
$256$ sites. Suitable boundary conditions have been chosen to obtain 
close-shell configurations in $|\Psi_{\rm MF}\rangle$.
 
\begin{figure}
\includegraphics[width=\columnwidth]{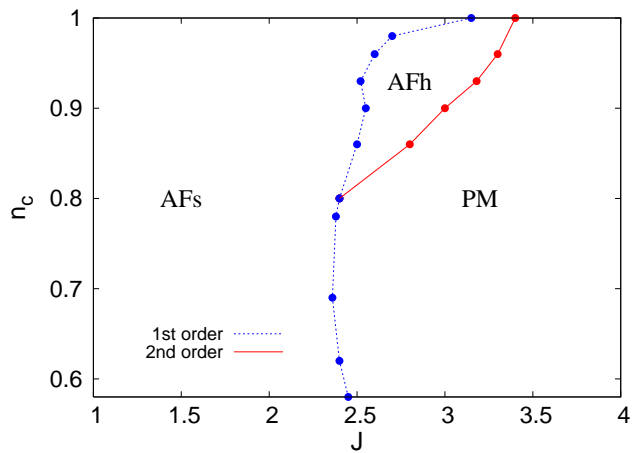}
\caption{\label{fig:phasemf}
(Color on-line) Mean-field phase diagram. AFs and AFh indicate 
antiferromagnetic phases with small and hole-like Fermi surfaces, respectively.
PM indicates a paramagnetic (metallic) phase. Continuous (dashed) lines mark
second-order (first-order) phase transitions.}
\end{figure}

\begin{figure}
\includegraphics[width=\columnwidth]{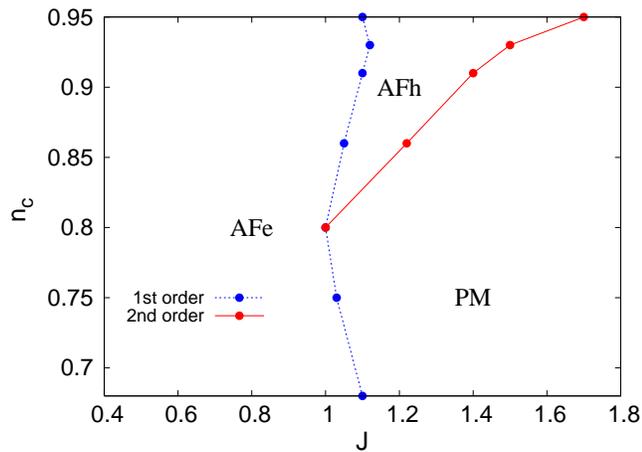}
\caption{\label{fig:phasevmc}
(Color on-line) Variational Monte Carlo phase diagram. The labels are the same
as in Fig.~\ref{fig:phasemf}. We note that, although qualitatively similar, the 
variational and mean-field phase diagrams are quantitatively quite different.}
\end{figure}

\begin{figure}
\includegraphics[width=\columnwidth]{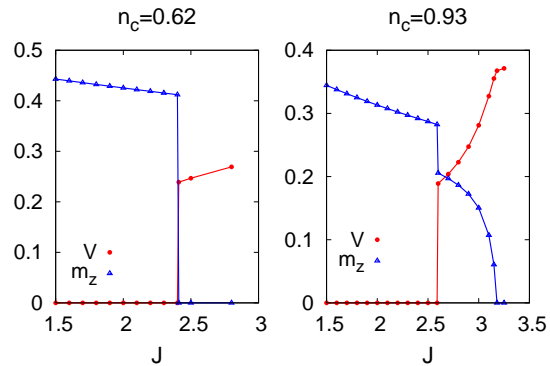}
\caption{\label{fig:mfparam}
(Color on-line) Mean-field order parameters (hybridization $V$ and total
magnetization $m_z=m_f-m_c$) as a function of the Kondo exchange for two 
different densities of the conducting electrons, $n_c=0.62$ (left panel) 
and $n_c=0.93$ (right panel).}
\end{figure}

\section{Results}\label{sec:results}

Here we present our numerical results on the KLM, first within the mean-field
approximation and then by the variational Monte Carlo approach.

\subsection{Mean-field results}\label{sec:MFresults}

The mean-field phase diagram, as a function of $J$ and the electron density 
$n_c$, is reported in Fig.~\ref{fig:phasemf} (for a direct comparison, we
report the variational Monte Carlo phase diagram in Fig.~\ref{fig:phasevmc}).
For $n_c>0.81$ we find two distinct phase transitions. When $J$ is small, 
the ground state has antiferromagnetic long-range order and displays a small 
Fermi surface, namely we obtain the AFs state. Here, the local $f$ electrons 
are totally decoupled from the conducting ones (the mean-field equations are 
solved by $V=0$) and do not contribute to the Fermi surface. This regime is 
dominated by the RKKY interaction that generates a magnetic pattern in the 
localized spins, and consequently also in the conducting ones: the 
magnetization of $f$ electrons is saturated, i.e., $m_f=0.5$, while $m_c$ is 
a smooth function, slightly increasing with $J$.

By increasing $J$, the Kondo mechanism becomes competitive with the RKKY 
interaction and we enter into another antiferromagnetic phase, where $c$ and 
$f$ electrons are hybridized. Here, there is a hole-like Fermi surface and, 
therefore, the phase is AFh. The hybridization $V$ has a finite jump at the 
transition, which is, therefore, first order. In Fig.~\ref{fig:mfparam}, we 
show the behavior of the hybridization $V$ and the total magnetization 
$m_z=m_f-m_c$. Eventually, by further increasing the local exchange, the Kondo
mechanism prevails and the system becomes a paramagnetic metal where conduction
electrons screen the local moments. The transition between the AFh phase and 
the paramagnetic metal is second order, with the magnetization that goes 
continuously to zero, see Fig.~\ref{fig:mfparam}. Moreover, the hybridization
is continuous though the transition and the topology of the Fermi surface 
does not change.

For smaller values of the conduction electron density, i.e., $n_c<0.81$,
the AFh state cannot be stabilized and there are only two phases: the AFs
for small Kondo exchange and the paramagnetic metal for large ones.
The phase transition between them is first order: both the antiferromagnetic
order parameter and the hybridization change abruptly from zero to a finite
value, see Fig.~\ref{fig:mfparam}. In this case, also the topology of the
Fermi surface changes across the transition.

\subsection{Variational Monte Carlo results}\label{sec:VMCresults}

Now, we turn to the variational Monte Carlo results, summarized in the phase 
diagram of Fig.~\ref{fig:phasevmc} (to be compared with the mean-field one, 
see Fig.~\ref{fig:phasemf}). Within the variational Monte Carlo, which 
improves substantially the Hartree-Fock energy, the $c{-}f$ hybridization 
parameter $V$ of the non-interacting auxiliary Hamiltonian 
${\cal H}_{\rm MF}$, is finite throughout the phase diagram.~\cite{ogata} 
It follows that the zero-temperature variational Fermi surface always includes
both $c$ and $f$ electrons. However, the optimized $V$ is tiny in the AFe 
phase and steps up discontinuously entering the AFh or PM phases. Therefore, 
a very small temperature can wash away the effects of $V$ in the AFe phase
(but not in the AFh and PM ones) thus better highlighting the differences 
between the phases. For this reason, we decided to follow 
Ref.~\onlinecite{lanata} and calculate, in the Brillouin zone, the emission
spectrum $A(k)$ of ${\cal H}_{\rm MF}$ at the chemical potential 
broadened with a low but finite temperature $T$: 
\begin{equation}\label{eq:Ak}
A(k) = - \int d\epsilon A(k,\epsilon) 
\frac{\partial f(\epsilon)}{\partial \epsilon},
\end{equation}
where $A(k,\epsilon)$ is:
\begin{equation}
A(k,\epsilon) = \sum_{n>0} \, |\langle n|c^\dagga_{k,\sigma}|0\rangle|^2
\delta(\epsilon - E_n + E_0),
\end{equation}  
where $|n\rangle$ are the unprojected (mean-field) states, with energies
$E_n$.

\begin{figure}
\includegraphics[width=\columnwidth]{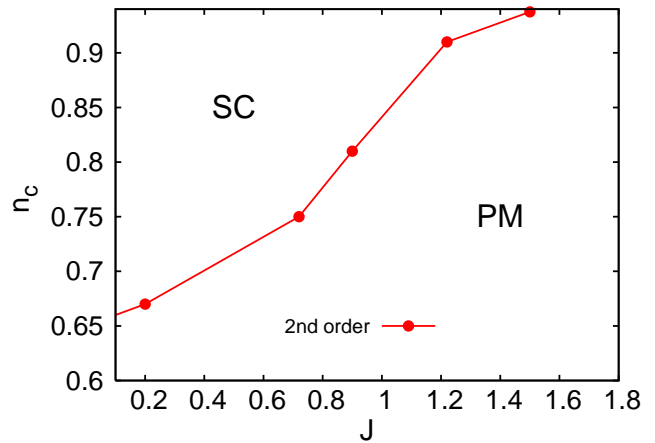}
\caption{\label{fig:phasepm}
(Color on-line) Variational phase diagram in the paramagnetic sector. 
PM and SC denote the paramagnetic metal and the $d$-wave superconducting state,
respectively. The transition between these two phases is always continuous.}
\end{figure}

\begin{figure}
\includegraphics[width=\columnwidth]{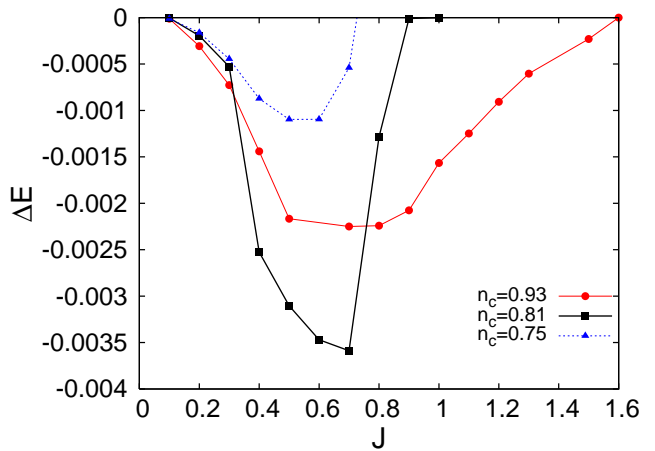}
\caption{\label{fig:energyBCS}
(Color on-line) Energy (per site) difference between the superconducting 
state and the metallic one as a function of $J$ for different values of the 
$c$-electron density.}
\end{figure}

We start our analysis by considering the paramagnetic sector, which is reacher
than the one obtained within mean-field approximation, and can shed some light
by disentangling Kondo effect from long-range magnetism. The paramagnetic 
phase diagram of the KLM, allowing for superconductivity, is shown in 
Fig.~\ref{fig:phasepm}. We find that, although (on-site or extended) $s$-wave 
pairing is never stabilized, a sizable $d$-wave pairing is obtained in a wide 
range of parameters, namely for $J \lesssim 1.5$ and $n_c \gtrsim 0.65$, and 
brings a non-negligible energy gain with respect to a normal phase. 
The {\it condensation energy} is reported in Fig.~\ref{fig:energyBCS} for 
three values of $n_c$. For $J \lesssim 0.1$, the pairing correlations of the 
unprojected state become very small, implying a tiny energy gain with respect 
to the normal state. We emphasize that superconductivity emerges only thanks 
to the electronic correlations brought by the Gutzwiller projector 
${\cal P}_f$, since pairing does not arise at the mean-field level. A finite 
$d$-wave pairing is thus generated by the antiferromagnetic $c{-}f$ exchange,
suggestive of similarities to analogous results found in $t{-}J$ models for 
cuprate superconductors.~\cite{dagotto,spanu} Indeed, as evident from 
Fig.~\ref{fig:energyBCS}, the condensation energy has a bell-like shape, 
with maximum at some intermediate values of $J$ and $n_c$. 

We mention that a very recent single-site dynamical mean-field theory (DMFT) 
calculation in the paramagnetic sector~\cite{bodensiek} finds evidence of 
$s$-wave superconductivity, whose maximum strength is reached, for a 
semicircular density of states, when $J\sim 1.6$ (translated in our units in 
which the bandwidth is $8t$) and $n_c \sim 0.86$, which we could not reproduce
by our simple variational wave function. 

\begin{figure}
\includegraphics[width=\columnwidth]{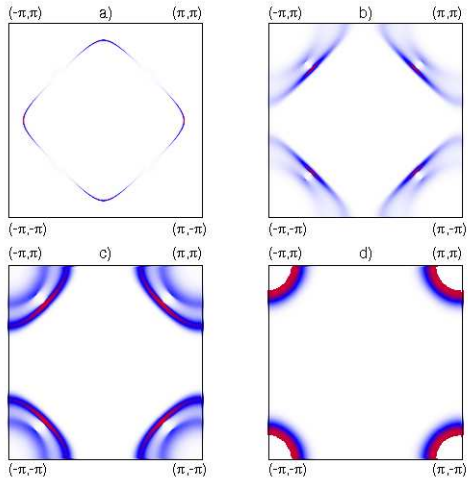}
\caption{\label{fig:DOS-PM}
(Color on-line) Emission spectrum $A(k)$, see Eq.~(\ref{eq:Ak}), broadened with 
a temperature $T=0.01$ for $n_c=0.91$. The top panels correspond to $J=0.2$ 
(left) and $J=0.8$ (right), the bottom one to $J=1.2$ (left) and $J=1.3$
(right). The lower three values of $J$ are inside the $d$-wave superconducting 
dome, while $J=1.3$ is already in the metallic phase. Note the change of 
topology as $J$ increases.}
\end{figure}

In Fig.~\ref{fig:DOS-PM}, we plot $A(k)$ for four different values of $J$, 
two well inside the superconducting region and two across the transition to 
the normal phase. Since the transition is continuous, the $T>0$ Fermi surface 
continuously change from electron-like to hole-like, see Fig.~\ref{fig:DOS-PM}.
A large spectral weight along the zone diagonals in the superconducting phase 
is observed whenever sizable pairing correlations are present, because of 
$d$-wave symmetry.    

\begin{figure}
\includegraphics[width=\columnwidth]{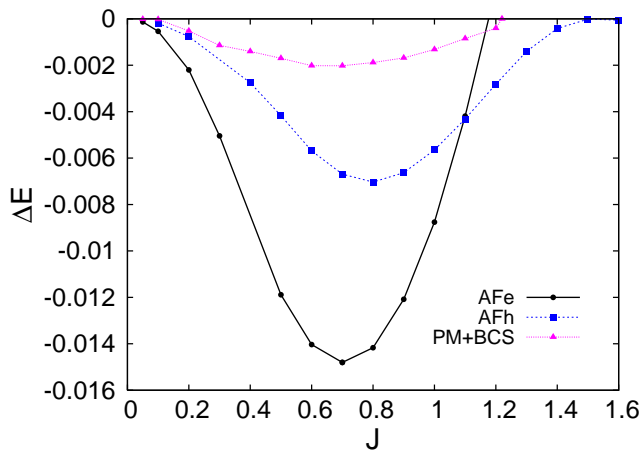}
\caption{\label{fig:energyAF}
(Color on-line) Energy (per site) difference between the antiferromagnetic, 
superconducting, and normal states as a function of $J$ for $n_c=0.91$.}
\end{figure}

\begin{figure}
\includegraphics[width=\columnwidth]{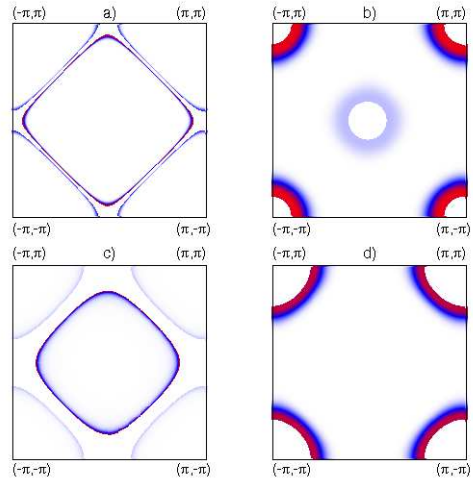}
\caption{\label{fig:DOS-AF} 
(Color on-line) Emission spectrum $A(k)$, see Eq.~(\ref{eq:Ak}), broadened with
a temperature $T=0.01$. Top left panel: $n_c=0.9375$ and $J=0.3$, inside the 
AFe phase. Top right panel: $n_c=0.9375$ and $J=1.3$, inside the AFh phase. 
Bottom left panel: $n_c=0.75$ and $J=0.2$, inside the AFe phase. 
Bottom right panel: $n_c=0.75$ and $J=1.2$, inside the PM phase. 
Note the shadow bands for the antiferromagnetic cases.}
\end{figure}

When we leave the paramagnetic sector and allow for antiferromagnetism, the 
latter prevails over superconductivity, which therefore disappear from the 
actual phase diagram, see Fig.~\ref{fig:phasevmc}. In other words, the energy 
gain of antiferromagnetism always overcomes that of superconductivity,
see Fig. \ref{fig:energyAF}, ruling out the possibility of a ground state 
with superconductivity and no magnetic order. This occurs at least in the 
bipartite nearest-neighbor hopping model that we have considered, where the 
only source of frustration is the conduction electron density $n_c$ lower than 
half-filling. We also investigated possible coexistence between 
antiferromagnetism and $d$-wave superconductivity, which we indeed found but 
only in the AFe region. However, we believe this result is only a finite size 
effect, since the energy gain by allowing $d$-wave pairing on top of magnetism
is tiny (at maximum, $\Delta E \simeq 10^{-4}t$) and, in addition, the size 
scaling of the actual order parameter (after Gutzwiller projection) suggests 
a vanishing value in the thermodynamic limit. We observe that the region of 
stability of the AFe phase is reduced substantially with respect to the 
corresponding AFs found at the mean-field level, compare 
Fig.~\ref{fig:phasevmc} with Fig.~\ref{fig:phasemf}, showing that the 
variational wave function can deal with Kondo screening better than mean field.

In Fig.~\ref{fig:DOS-AF} we draw $A(k)$ for different values of $n_c$ and $J$.
The left panels are inside the AFe phase, and show a spectral distribution 
at the chemical potential that corresponds to a small, electron-like, Fermi 
surface. On the contrary, the right panels (the top one inside the AFh phase 
and the bottom one in the PM region) indicate a larger Fermi surface that 
still contains $f$ electrons at that value of temperature $T$. We note the 
signals of shadow bands in the antiferromagnetic $A(k)$ of the top panels and 
left bottom one. 

We finally mention that the phase diagram of Fig.~\ref{fig:phasevmc} agrees 
pretty well with that obtained by Watanabe and Ogata,~\cite{ogata} who also 
use a variational wave function similar to ours, though with less variational
freedom since it allows only $c{-}f$ hybridization and antiferromagnetism. 
Indeed, we find that all additional variational parameters, e.g., $f{-}f$ 
hopping and superconductivity, do not change appreciably the energy, hence 
the phase diagram.  Moreover, the variational phase diagram bears similarity 
also to that one obtained by Martin, Bercx, and Assaad~\cite{assaad} by the 
dynamical cluster approximation, although in the latter case all transitions 
seem continuous.  

\section{Conclusions}\label{sec:conc}
 
In conclusion, we have studied, by mean-field and variational Monte Carlo 
techniques, the Kondo lattice model on a square lattice. The mean-field phase 
diagram is qualitatively but not quantitatively similar to the variational 
Monte Carlo one. Restricting the analysis to the paramagnetic sector, we have 
found by variational Monte Carlo a large region of $d$-wave superconductivity, 
which however disappears from the phase digram once we allow for 
antiferromagnetism. It is well possible that, if magnetic frustration is added
besides that due to $n_c<1$, superconductivity could emerge and intrude 
between the antiferromagnetic phase and the paramagnetic one, thus offering a 
possible explanation to what is observed in many heavy fermion compounds. 

The variational Monte Carlo phase diagram is practically the same as that 
obtained by Watanabe and Ogata~\cite{ogata} by a similar technique. 
In particular we also find that the onset of antiferromagnetism and the 
breakdown of Kondo effect are not simultaneous close to compensation, $n_c=1$. 
Moreover, we find that the Kondo collapse always occurs via a first order 
phase transition and is accompanied by a redistribution of low-temperature 
spectral weight at the chemical potential inside the Brillouin zone. 
 
\acknowledgments
 
We would like to thank H. Watanabe for providing us with his data for 
preliminary check of our results. This work has been supported by PRIN/COFIN 
2010LLKJBX\_004.

\end{document}